\documentclass[final,3p]{elsarticle}
\usepackage{latexsym}
\usepackage{amssymb, amsmath, bm, physics}
\usepackage{subcaption}
\usepackage{mathptmx}
\usepackage{aas_macros}
\usepackage{graphicx}

\journal{Annals of Physics}

\begin{document}

\begin{frontmatter}

\title{Intrinsic quantum dynamics of particles in brane gravity}
\author{Shahram Jalalzadeh}
\address{Departamento de F\'{i}sica, Universidade Federal de Pernambuco, Recife, PE,  50670-901, Brazil}
\ead{shahram.jalalzadeh@ufpe.br}

\begin{abstract}
The Newtonian dynamics of particles in brane gravity is investigated. Due to the coupling of the particles' energy-momentum tensor to the tension of the brane, the particle is semi-confined and oscillates along the extra dimension. We demonstrate that the frequency of these oscillations is proportional to the kinetic energy of the particle in the brane.
We show that the classical stability of particle trajectories on the brane gives us the Bohr--Sommerfeld quantization condition. The particle's motion along the extra dimension allows us to formulate a geometrical version of the uncertainty principle. Furthermore, we exhibited that the particle's motion along the extra dimension is identical to the time-independent Schr\"odinger equation. The dynamics of a free particle, particles in a box, a harmonic oscillator, a bouncing particle, and tunneling are re-examined. We show that the particle's motion along the extra dimension yields a quantized energy spectrum for bound states.
\end{abstract}


\end{frontmatter}



\section{Introduction}\label{S1}

For a long time, physicists have sought to reconcile quantum physics with gravity. Following the failure of gravity quantization, several physicists argued that gravity is an induced (or emerged) rather than a fundamental force and that quantization is thus superfluous (see, for example, \cite{qi2018does,Steinacker,Steinacker:2010rh,Barcelo:2005fc} and references therein). Sakharov's induced gravity theory \cite{2000GReGr} is possibly the most known of these alternate theories. Induced gravity theories desire to show how specific underlying infinitesimal degrees of freedom, approximated by a mean field, lead to the formation of curved space-time. This method is believed to be similar to the one used to produce fluid dynamics by Bose--Einstein condensation. Another explanation for the emergence of gravity is that non-geometric statistical rules are responsible for the existence of curved space-time \cite{Jacobson:1995ab,Verlinde:2010hp}.  Such models have so far proven successful in reproducing some of the potential features of quantized general relativity, but a complete picture is still missing.

In distinction to the mentioned group, others have proposed that  quantum mechanics (QM) is not primary and that something more fundamental lies beneath it. To be as provocative as possible, we will point out that QM's fundamental equations (the Schr\"odinger equation (SE) and its relativistic expansions) are all `empiric' laws \cite{reichenbach}. In this sense, the term `empiric' has a distinct connotation since it alludes to the fact that particular laws cannot be drawn from basic principles \cite{reichenbach}. This is despite the fact that previous attempts to derive these equations from other assumptions, such as the wave-particle analogy, failed.
The arguments of Dirac \cite{dirac1981principles}, who attempted to derive his equation from the challenge of describing microsystems using equations that contain the second derivative concerning time, are a remarkable illustration of this. Even though Dirac's assumption resulted in a successful equation (the Dirac equation for particles with spin $1/2$), several second-order equations with respect to time are known to describe satisfactorily massive particles with integer spin: the Klein--Gordon and Proca--Yukawa equations are examples. Many attempts have been made to infer the quantum equations from `first' principles, as the above-mentioned equations must be regarded as empiric rules. 
As an example of these `whole theories,' we can mention the Feyn\`es--Nelson stochastic formulation \cite{1952ZPhy13281F,PhysRev.150.1079,1997PhLA,Novello:2009tn,2003CSF},  scale relativity \cite{Nottale:2007ne}, Weylian geometry \cite{1936PPS33F,1984JMP25.2477S,1992FoPh569C,1993FoPhL207W,1984PhRvD216S,Godfrey:1984nc,Novello:2009tn,Falciano:2010zz,PhysRevD.41.431}, Finsler geometry \cite{2019IJGMM0098B,2016AnPhy.239T,Liang:2019yqy}, and Kaluza--Klein (KK) theory.

In this manner, the extra dimension of KK theory was connected to London's spin of the electron \cite{Ernst}.
Also, the idea of relating Schr\"odinger's wave function with the extra metrical component $g_{55}$  of KK was put to work. Gonseth and Juvet, in their paper \cite{Gonseth}, declared that the field equations of KK theory give the laws for the gravitational field, the electromagnetic field, the laws of motion of a charged particle, and the Schr\"odinger wave equation. In this manner, they provided a framework in which to take the gravitational and electromagnetic laws and make it possible for quantum theory to be included. Also, W. Wilson \cite{doi:10.10} and H.T. Flint \cite{1930RSPSA.126.644F,1958NCim680F} obtained the SE in the spirit of KK theory. 

The next step in driving SE from extra dimension comes from P.S. Wesson's works \cite{Wesson1,Wesson2,Wesson3}, which developed the above ideas to the modern interpretation of the KK theory with a non-compact extra dimension.
The above examples show that the attempt to understand the hidden foundations of quantum mechanics in the geometry of space-time and gravity is as old as QM itself.

 Despite the conceptual appeal of the Feyn\`es--Nelson stochastic formulation \cite{1952ZPhy13281F,PhysRev.150.1079}, their model caused more problems than they solved, and they fell short of their primary purpose of shedding light on the physical mechanism that underpins matter's wave-like properties. In truth, it includes an uncommon mechanism that causes the individual to have a `mysterious dependence on the statistical ensemble of which it is a member.' This feedback mechanism has no counterpart in classical (Newtonian) mechanics or stochastic process theory, and its foundation is completely quantum. Therefore, it seems that the `classical' formulations of  Feyn\'es--Nelson is more about appearance than content. 
 
 However, one example exists in which a statistical ensemble's interplay with its members is a natural consequence of fundamental principles, and this example comes from general relativity. Consider, for example, Synge's statistical ensemble of freely falling test particles \cite{1960rgtbookS}. The ensemble scalar density determines the geodesic followed by each particle using Einstein field equations. Because some geometrical fields may provide the requisite feedback between the ensemble and its members, QM as a classical theory may have the same logical structure as general relativity \cite{Santamato1986}.
In articles, \cite{2009EL8740006R,2007GReGr..39.1467M}, it has been shown that the extrinsic curvature of the 3-brane may account for quantum phenomena. In this scheme, quantum and gravitational phenomena are taken on the same logical grounds: gravity is associated with the metric tensor, and quantum effects are associated with the second fundamental form, i.e., the notion of deformation of 3-brane along the extra dimension. The first and second fundamental forms are logically independent geometric entities so that we may look at the problem of a quantum particle in an arbitrarily gravitational field.

Within the context conveyed by the above paragraphs, this paper aims to introduce a new perspective on the classical description of wave mechanics. We propose bringing to the discussion a different perspective, broadening the scope of discussion using the braneworld framework. Hence, the rest of this paper is organized as follows. Section 2 introduces the dynamics of non-relativistic particles in the braneworld scenario. Then, in section 3, we extend the results of Ref. \cite{2009EL8740006R} to a general Shiromizu, Maeda, and Sasaki (SMS) braneworld scenario: for a closed trajectory of a particle in a 3-brane, the orbit is stable if it satisfies the Bohr--Sommerfeld quantization role. Section 4 solves the equations of motion for a free particle, the infinite square well, the linear harmonic oscillator, and the bouncing particle. We show that a free test particle oscillates along the extra dimension. The particle appears in the brane periodically with an oscillation wavelength equal to the de Broglie wavelength. The particle's energy is quantized for the infinite square well, bouncing particle, and the harmonic oscillator, the same as the QM. Section 5 shows that the particle's equation of motion along the extra dimension is equivalent to the SE. Indeed, we reanalyze the tunneling through a potential barrier. Section 6 contains our conclusions and a thorough discussion. For readers who are interested in our more detailed calculations, we have also included an appendix.

\section{Newtonian dynamics of particles}

According to the general formulation of SMS braneworld scenario \cite{2000PhRvD62b4012S}, our $4D$ universe (the brane) is embedded locally and isometrically in a codimension-1 ambient (the bulk) space, in which the standard model fields are confined to the brane, while gravity propagates in all spatial dimensions.  
If we denote the extra dimension with coordinate $\psi$, then our brane is located in the position $\psi=0$ and in the Gaussian normal coordinates $x^A:=(x^\mu,\psi)$\footnote{Capital letters run from 0 to 4, whereas Greek letters run from 0 to 3.} the line element of the bulk is
\begin{eqnarray}\label{1a}
ds^2=g_{\mu\nu}(x^\alpha,\psi)dx^\mu dx^\nu+d\psi^2.
\end{eqnarray}
Using Israel's junction conditions, the extrinsic curvature of the brane, $K_{\mu\nu}(x^\alpha,0)$, may be replaced by the energy-momentum tensor of the brane. We may replace the jump in the extrinsic curvature by twice the extrinsic curvature's value at the brane location if we assume mirror symmetry, or $\mathbb Z_2$-symmetry, throughout the brane \cite{2000PhRvD62b4012S}
\begin{eqnarray}\label{1b}
K_{\mu\nu}(x^\alpha,0)=-\frac{\kappa^2_{(5)}}{2}\left\{T_{\mu\nu}-\frac{1}{3}(T-\lambda)g_{\mu\nu}(x^\alpha,0)\right\},
\end{eqnarray}
where $\kappa^2_{(5)}$ is Einstein’s gravitational
constant of the bulk, $g_{\mu\nu}(x^\alpha,0)$ is the metric of the brane, $T_{\mu\nu}$ is the energy-momentum tensor of the confined fields and matters, $T:=g^{\mu\nu}(x^\alpha,0)T_{\mu\nu}$ is its trace, and $\lambda$ is the tension of the brane in the bulk space. In addition, the $4D$ Newton's gravitational constant, $G_\text{N}$, and the $4D$ cosmological constant, $\Lambda$, are given by  
\begin{eqnarray}\label{1c}
8\pi G_\text{N}=\frac{1}{6}\kappa_{(5)}^4c^4\lambda,~~~\Lambda=\frac{\kappa_{(5)}^2}{2}\left(\Lambda_\text{B}+\frac{\kappa_{(5)}^2}{6}\lambda^2\right),
\end{eqnarray}
respectively, where $\Lambda_\text{B}$ represents the cosmological constant of the bulk space \cite{2000PhRvD62b4012S}.

Here, we assume that while gauge fields and macroscopic matter are confined to the brane, a microscopic particle with mass $m$ and electric charge $q$ can have classical oscillations along the extra dimension throughout the brane. When a particle oscillates along the extra dimension, we call it a semi-confined particle. To obtain the equations of motion of such a particle, regarding the line element (\ref{1a}),  we start with a $5D$ action for the particle \cite{G1,G2,G3,G4,G5,G6,G7,G8,G9,G10}
\begin{equation}\label{1d}
S=-\displaystyle\int\left\{mc\sqrt{-g_{\mu\nu}(x,\psi)\dot x^\mu\dot x^\nu-\dot \psi^2}+\frac{q}{c}A_\mu(x)\dot x^\mu\right\}d\tau,
\end{equation}
where $\tau$ is a $5D$ affine parameter, $A_\mu(x^\alpha)$ is the confined electromagnetic 4-potential, and  an overdot represents derivative with respect to $\tau$. The variation of the above action gives us the equations of motion \cite{G1,G2,G3,G4,G5,G6,G7,G8,G9,G10}
\begin{equation}\label{1e}
\begin{split}
\ddot x^\mu&+\Gamma^\mu_{\alpha\beta}\dot x^\alpha \dot x^\beta=-g^{\alpha\mu}{\partial_\psi g_{\alpha\nu}}\dot x^\nu\dot \psi+\frac{q}{mc}g^{\mu\nu}F_{\nu\beta}\dot x^\beta,\\
\ddot{\psi}&-\frac{1}{2}\partial_\psi g_{\mu\nu}(x,\psi)\dot x^\mu\dot x^\nu=0,
\end{split}
\end{equation}
where $\Gamma^\mu_{\alpha\beta}$ are the affine connection components, and $F_{\mu\nu}:=A_{\nu,\mu}-A_{\mu,\nu}$ is the electromagnetic tensor. To obtain the non-relativistic limit of the Eqs.(\ref{1e}), we follow the following steps: First, we expand the metric of the bulk space, $g_{\mu\nu}(x^\alpha,\psi)$, in the vicinity of the brane. One can show that \cite{ovalle2020beyond}
\begin{equation}
\label{1f}
g_{\mu\nu}(x^\alpha,\psi)=g_{\mu\nu}(x^\alpha,0)-2K_{\mu\nu}\psi+\Big\{K_{\mu\beta}K^\beta_{~~\nu}-\mathcal E_{\mu\nu}-\frac{1}{6}\Lambda_\text{B}g_{\mu\nu}(x^\alpha,0)\Big\}\psi^2+\mathcal O(\psi^3),
\end{equation}
where $g_{\mu\nu}(x^\alpha,0)$ is the metric of the brane, $\mathcal E_{\mu\nu}$ is the projection of the Weyl tensor of the bulk space on the brane,
and $K_{\mu\nu}$ is given by junction condition (\ref{1b}). $\mathcal E_{\mu\nu}$ is a constituent of the $5D$ Weyl tensor that contains information about the gravitational field outside the brane. If the bulk space-time is not entirely de Sitter (or anti de Sitter), it is non-vanishing.

In the second step, we split the energy-momentum tensor of confined matter fields in junction condition (\ref{1b}) into the energy-momentum of the particle itself, its interaction with fields,  and all other confined fields. i.e., $T_{\mu\nu}(\text{particle})+T_{\mu\nu}(\text{fields})+$$T_{\mu\nu}(\text{interaction})$. 

Third, for simplicity, we assume that the bulk space is purely de Sitter with a small $5D$ cosmological constant $\Lambda_\text{B}$. As a result, $\mathcal E_{\mu\nu}$ vanishes, and $\Lambda_\text{B}$ is expected to be
negligible in the non-relativistic limit. As previously stated, the purpose of this work is to investigate the influence of extra dimension on microscopic systems, such as an atom. The influence of minor gravitational corrections by the bulk cosmological constant on Newton's law of gravity is insignificant in these situations. As a result, $\Lambda_\text{B}$ can be removed; for additional information, see \ref{App}.


Besides, in the dynamics of microscopic and non-relativistic systems, the effect of distance matter (dark energy, distribution of matter in the Universe, etc.) is negligible, i.g., $T_{\mu\nu}(\text{rest of the Universe})=0$. Smallness of $\Lambda_\text{B}$ and the vanishing $\mathcal E_{\mu\nu}$ in Eqs.(\ref{1c}) gives us $\lambda=\frac{\Lambda c^4}{4\pi G_N}$. Inserting the values of the constants in the right-hand side of this equation (see (\ref{value1}) and (\ref{value2}) for the measured values of these constants), we find $\lambda=6.6~ \text{GeV}/\text{m}^3$ which is negligible respect to the energy-momentum tensor of the particle in Eq.(\ref{1b}).

At the zero approximation (which means, $\psi=\dot\psi=0$), the energy-momentum of the particle with mass $m$ and proper-time $d\bar\tau$ \footnote{All quantities with bar symbols denote restricting to the brane. For example, $d\bar\tau$ represents an affine parameter in 3-brane. } is
\begin{equation}\label{T11}
T_{\mu\nu}=\frac{m}{\sqrt{-g(x,0)}}\frac{dx^\mu}{d\bar\tau}\frac{dx^\nu}{d\bar\tau}\frac{d\bar\tau}{dt}\delta^3(\vec x-\vec x(\bar\tau(t)).
\end{equation}

It should be noted that the relation (\ref{T11}) does not reflect the energy-momentum tensor of a semi-confined particle. Regarding the geodesic equations (\ref{1d}), the particle has a velocity component along the extra dimension, and as we will see later, it is semi-confined to the brane. Let us look at the meaning of the preceding equation. In general relativity, matter curves space-time intrinsically. In the presence of the extra dimension, it also bends space-time extrinsically.
In the same way, the presence of a particle modifies the 3-brane extrinsically via Eq.(\ref{1b}). As we linearized the geodesic equations up to the second order of $\psi$, the bent of the 3-brane is related to the energy-momentum tensor of a point-like particle given by Eq.(\ref{T11}) up to the first order of the perturbation. In other words, up to the first order of perturbation of tensor quantities, by putting the energy-momentum tensor as mentioned earlier into Eq. (\ref{1b}), one may get the extrinsic curvature of the 3-brane at the particle's position on the brane (see Fig. \ref{fig00} in \ref{App}). Let us go through some specifics. The existence of a particle in the vicinity, $U_x$, of the brane alters the intrinsic and extrinsic features of the brane.
 While the intrinsic deformation of the 3-brane for a microscopic particle may be neglected, as demonstrated in \ref{App}, the extrinsic deformation of bulk properties is substantial.
Extrinsic curvature offers a measure of the deviation from the brane and its tangent plane at every point of the 3-brane, which is referred to as geometry bending \cite{Maia:2001gq}.
Replacing the above energy-momentum tensor into (\ref{1b}) indicates that the particle energy and momentum bend the space-time manifold along the extra dimension locally at the particle's position. 
Therefore, it simply provides the energy required to compute the bending of space-time in bulk space at the location of the semi-confined particle.

As we know, we can not define a `coherent product' of distributions. Therefore, as a result of the existence of a Dirac delta distribution in Eq.(\ref{T11}), the expansion (\ref{1f}) is not well-defined. Thus, we approximate the delta function in the energy-momentum tensor of the particle with the inverse of the classical volume of the electron
\begin{equation}
\delta^3\left(\vec x-\vec x(\bar\tau(t)\right)\simeq
\begin{cases}
0,\hspace{1.5cm}r>\beta r_e,\\
\frac{1}{\frac{4\pi}{3}(r_e\beta)^3},~~~~~~r\leq\beta r_e,
\end{cases}
\end{equation}
where $\beta=\pi/2$, $r_e=\frac{e^2}{4\pi \varepsilon_0m_ec^2}$ is the classical radius of the electron, $m_e$, is the electron's mass, and $e$, is the charge of the electron. Note that this volume does not represent the actual, measurable volume of the electron.
One may believe that, under the appropriate circumstances, even a macroscopic object can be idealized as a point-like particle and that the Newtonian limit of our geodesic equations, particularly along the extra dimension, can correctly explain the motion of such a particle. As a result, it appears quite arbitrary and unnatural that the classical electron volume was chosen to support the delta of anybody. The particle's kinetic energy is actually quite crucial in this circumstance, as shown in \ref{App}: Because a macroscopic particle has a large mass, its oscillation along the extra dimension is fairly limited. The amplitude of its oscillations is actually smaller than the reduced Compton wavelength. Furthermore, the angular frequency of the oscillation (see Eq.(\ref{1g:b})), which determines how frequently it oscillates, is proportional to its kinetic energy. As a result, a macroscopic object would not feel the extra dimension, and it is most likely restricted to the brane. 
On the other hand, the electron is the lightest subatomic particle\footnote{The neutrino situation is comparable to the graviton case. As mentioned in \ref{App}, gravitons can spread across bulk space until they approach the bulk de Sitter horizon.} and the corresponding extrinsic radius is larger than other subatomic particles. By Choosing the classical electron volume as the regulator of the Dirac delta distribution, all other subatomic particles  will oscillate inside the same extrinsic curvature bubble. If we choose a heavier particle volume,  the electron becomes an exception, and the equations of motion cannot explain it.

These assumptions lead to obtaining the non-relativistic limit of energy-momentum tensor $T_{\alpha\beta}$, and via junction condition (\ref{1b}), the components of extrinsic curvature of the brane in the vicinity of the particle in the brane
\begin{equation}\label{non}
    \begin{split}
 &T^{00}=\rho\bar E_k,~~~~~~T^{ij}=\rho m\bar v^i\bar v^j,~~~~~T^{0i}=0,\\
 &K^{00}=\frac{2}{3}\kappa_{(5)}^2\rho\bar E_k,~~~~~~K^{0i}=0,~~~~~~ K^{ij}=\frac{\rho\kappa_{(5)}^2}{2}(m\bar v^i\bar v^j-\frac{1}{3}\delta^{ij}\bar E_k),\\
& K_{\alpha\gamma}K^\gamma_{\,\,\beta}\frac{dx^\alpha}{d\bar\tau}\frac{dx^\beta}{d\bar\tau}=-\frac{4}{9}\kappa^4_{(5)}\rho^2c^2\bar E_k^2,~~~~~~ K_{\alpha\beta}\frac{dx^\alpha}{d\bar\tau}\frac{dx^\beta}{d\bar\tau}=0,
    \end{split}
\end{equation}
where $\rho=1/\frac{4\pi}{3}(\frac{e^2\beta}{4\pi \varepsilon_0m_ec^2})^3$, and $\bar E_k=\frac{m}{2}(\bar v_i)^2$ represents the kinetic energy of the particle at the zero order of the perturbation, i.e., $(\psi=d\psi/dt=0,~\bar v^i=dx^i/dt)$.
Now, to obtain the non-relativistic limit of the geodesic equations (\ref{1e}), we employ Cartesian coordinates and require the following conditions: (i) the gravitational field is weak so that we can write $g_{\mu\nu}(x^\alpha,0)$ as
the Minkowski metric plus a small correction, $h_{ij}$, (ii) the gravitational field is stationary,
i.e., independent of time, and (iii) the motion of particles is non-relativistic. Also, to obtain the space component of using (\ref{1f}), we simplify
\begin{equation} \label{111a}
       \Gamma^i_{\alpha\beta}\dot x^\alpha\dot x^\beta=-\frac{1}{2}g^{ij}\partial_jg_{00}=c^2\partial_i\left(-1+ h_{00}-\frac{4}{c^2\hslash^2}\bar E_k^2\psi^2\right)
    =\partial^i\Phi(x^j)-\frac{4}{\hslash^2}\psi^2\bar E_k\partial^i\bar E_{k},
\end{equation}
where in the third equality, we used the seventh equation obtained in (\ref{non}), $\Phi(x^j)$ represents a weak gravitational field defined by $h_{00}=-2\Phi(x^j)/c^2$, and $\hslash$ is
\begin{equation}
    \label{hslash}
    \hslash:=\frac{3}{c\rho\kappa_{(5)}^2}.
\end{equation}
The second term in (\ref{111a}) is the second order with respect to $\Psi$. Therefore, one can use the zero-order (the equations of motion of a particle which is totally confined) equations of motion to obtain $\partial^i\bar E_k=-md\bar v^i/dt$. By substituting  (\ref{111a}) into the first geodesic equation in (\ref{1e}), it is easy to show that its Newtonian limit for space component of 4-velocity, $\dot x^\mu$, is given by 
\begin{equation}
    \label{geo1a}
   \frac{d^2x^i}{dt^2}+\partial^i\Phi+\frac{4}{\hslash^2}\psi^2\frac{d^2x^i}{dt^2}=\frac{q}{mc}\delta^{ij}F_{jk}\frac{dx^k}{dt}+\frac{q}{m}\delta^{ij}F_{j0}.
\end{equation}
The electric and magnetic fields can be written in terms of the Faraday tensor  $F_{io}=E_i,~~F_{ij}=\varepsilon_{ijk}B^j$. Using these relations, the above equation of motion simplifies to
 \begin{equation}\label{1g:a}
 \boxed{ M  \frac{d^2x^i}{dt^2}=q\left(E^i+(\vec v\times\vec B)^i\right)   - \partial^iV(x),}
    \end{equation}
    where
\begin{equation}
    \label{sshh}
    M:=\left( 1-\frac{4m}{\hslash^2}\bar E_k\psi^2\right)m,
\end{equation}
is the `effective' mass of the particle, $E^i$ is the electric field, $B^i$ is the magnetic field, $V(x)$ is the static external potential energy (including weak gravitational potential $\Phi$ and emerged macroscopic potentials from the electric fields, for example, harmonic oscillator potential or finite-infinite potential box), $v^i=dx^i/dt$ is the Newtonian 3-velocity. 

The constant $\hslash$ in (\ref{hslash}) is defined in terms of the energy scale of the bulk space. Let us rewrite it in terms of $4D$ fundamental constants. Using Eqs.(\ref{1c}) we find $\kappa_{(5)}^2=\frac{8\pi G_N}{c^4}\sqrt{\frac{3}{\Lambda}}$. Inserting this value and the value of $\rho$ defined in Eq.(\ref{non}) into the definition of  $\hslash$ in (\ref{hslash}) gives us
\begin{equation}\label{1h}
 \boxed{   \hslash=\frac{1}{2G_Nm_e^3c^3}\sqrt{\frac{\Lambda}{3}}\left(\frac{\beta e^2}{4\pi\varepsilon_0} \right)^3.}
\end{equation}

In addition, one can obtain the Newtonian limit of the second geodesic equation in (\ref{1e})
    \begin{equation}
\boxed{   \sqrt{\bar E_k}\frac{d}{dt}\left(\frac{1}{\sqrt{\bar E_k}}\frac{d\psi}{dt} \right)+\frac{4}{\hslash^2}\bar E_k^2\psi=0.}\label{1g:b}
    \end{equation}
As Eq.(\ref{1g:a}) shows, the particle's motion along the extra dimension affects the particle's mass. Thus, for a closed trajectory in brane, see Fig.\ref{fig0}, the trajectory of the particle will be stable {\it if the initial and final values of the extra dimension in one loop are equal}. 

The total energy of the particle is
  \begin{equation}\label{Energy}
      E=\frac{1}{2}m\bar v^2+V(x)+\Phi(x)+\mathcal O\left(\psi^2,\left(\frac{d\psi}{dt}\right)^2\right),
  \end{equation}
where $\bar v^2:=\delta_{ij}\bar v^i\bar v^j$, and $\Phi(x)$ is the electric potential energy. By inserting (\ref{1g:a}) and (\ref{1g:b}) into $dE/dt$ obtained from (\ref{Energy}), it is easy to verify that $E$ (up to $\mathcal O(\psi^2)$) is conserved quantity.

Let us calculate the numerical value of the constant $\hbar$ defined in relation (\ref{1h}). If we insert the values \cite{523381} of the following constants
\begin{equation}\label{value1}
\begin{split}
   G_N&=6.67430(15)\times10^{-11}~\text{m$^3$/Kg$\cdot$s$^2$}, \\
   \frac{1}{4\pi\varepsilon_0}&=8.98755517923(14)\times10^9~\text{N$\cdot$ m$^2$/C$^2$},\\
   m_e&=9.1093837015(28)\times10^{-19}~ \text{Kg},\\
   c&=2.99792458\times10^8~\text{m/s},\\
   \beta&=\frac{\pi}{2},~~~
   e=1.602176634\times10^{-19}~\text{C},
   \end{split}
\end{equation}
  and the value of the cosmological constant from the Planck Collaboration (2018)	 \cite{Planck:2018vyg} 
  \begin{equation}\label{value2}
      \Lambda=1.088(30)\times10^{-52}~\text{1/m$^2$},
  \end{equation}
   into Eq.(\ref{1h}) we obtain
   \begin{equation}
       \hslash=1.054(015)\times10^{-34}~\text{J}\cdot \text{s}.
   \end{equation}
   This agrees with the experimental value of the reduced Planck's constant $\hslash_\text{exp}=1.054571817\times10^{-34}~J\cdot s$. To obtain a more precise result, we must wait for more accurate measurements of the gravitational and cosmological constants.

Before we go any further, let us draw the reader's attention to a few key points. First and foremost, equations (\ref{1g:a}) and (\ref{1g:b}) are our primary tools in the remainder of this paper. Also, $\psi$ in equation (\ref{1g:b}) represents the particle's coordinate along the extra dimension. $\psi$ is commonly used as the wave function. We represent the extra coordinate with $\psi$ because it is related to the wave function in QM, as we will see in the section \ref{Schro}.

\section{The Bohr--Sommerfeld quantization condition and Heisenberg inequality}
Before investigating specific examples, let us see how the equations of motion (\ref{1g:a}) and (\ref{1g:b}) give us the Bohr--Sommerfeld quantization rule. As observers, fields, and macroscopic measurement apparatuses are entirely confined to the 3-brane, there are some hidden variables in these equations: Eq.(\ref{1g:b}) shows that the particle oscillates around the 3-brane located at $\psi=0$ and most of the time it is in the bulk space and observer does not `see' it. To `observe' the particle, we should pull it back into the 3-brane at the moment of observation, and as these two equations of motion show, the action of observing disturbs the momentum and the position of the particle. Therefore, one can extrapolate that the initial position and velocity of the particle, $(x^i_0,p^i_{0})$ on the brane, are the hidden variables of the theory. Let us consider the particle moves under the influence of a central electric field (Coulomb problem) or a constant magnetic field in which the corresponding trajectory in the brane is a closed stable loop. A general solution of (\ref{1g:b}) can be expressed in the form 
\begin{equation}
    \psi=R\sin(S).
\end{equation}
 Substituting this solution into (\ref{1g:b}) gives us
\begin{equation}
 \label{Melin1a}
      S=C\oint \frac{\sqrt{\bar E_k}}{R^2}dt-\delta,
\end{equation} 
and
\begin{equation}      \label{Melin1b}
     \sqrt{\bar E_k}\frac{d}{dt}\left( \frac{1}{\sqrt{\bar E_k}}\frac{dR}{dt}\right)+\frac{4\bar E_k^2}{\hslash^2}R=\frac{C^2\bar E_k}{R^3},
\end{equation}
where $C$ and $\delta$ are arbitrary constants.
One can find an approximate solution of (\ref{Melin1b}) similar to the `semi-classical approximation' in QM. If we assume $\ddot R/R\ll \frac{4\bar E_k^2}{\hslash^2}$ then one can find
\begin{equation}\label{1i}
    \psi(t)=\frac{N}{(\bar E_k)^\frac{1}{4}}\sin(\frac{2}{\hslash}\int_0^t\bar E_kdt-\delta),
\end{equation}
where $N$ is a constant.

\begin{figure}[ht]
\centering
\includegraphics[width=10cm]{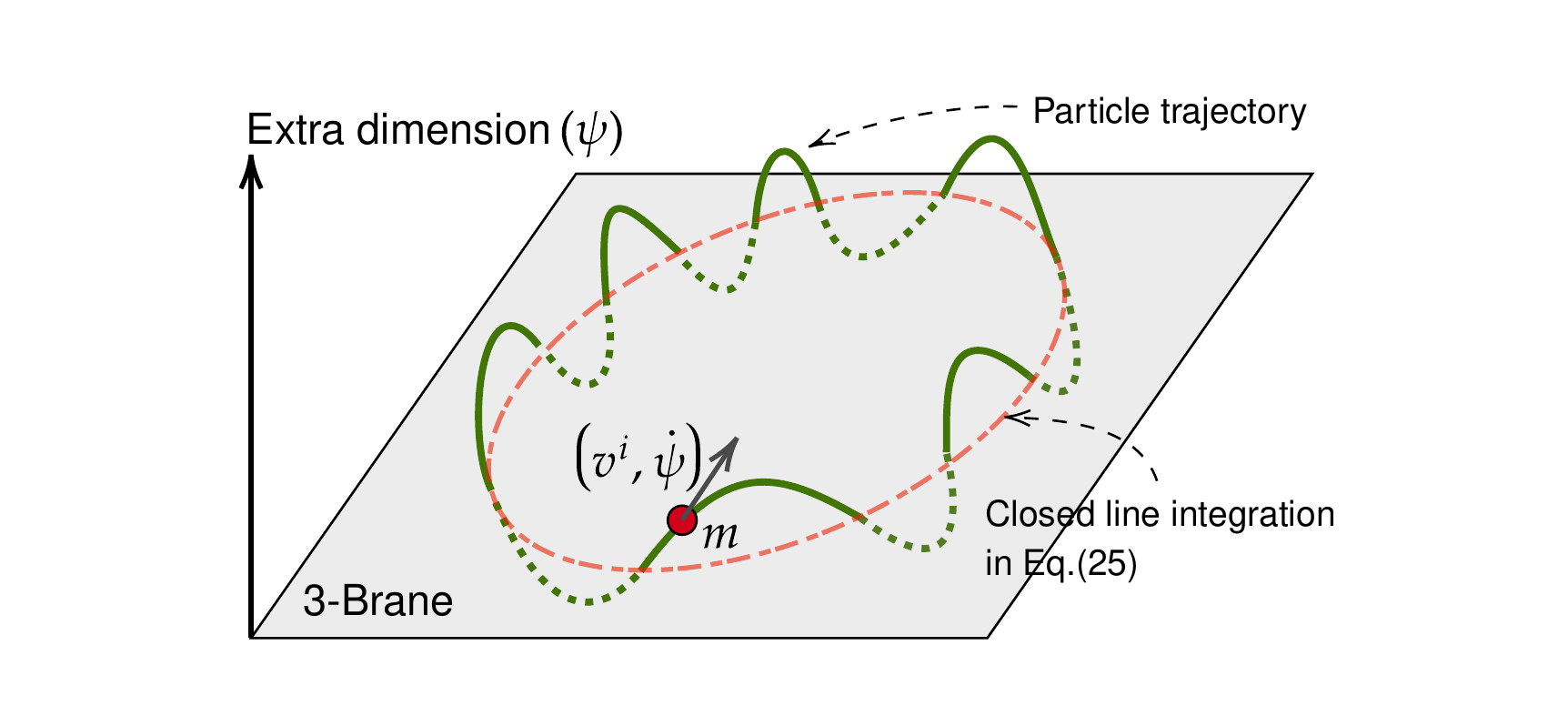}
\caption{A trajectory of a particle in the bulk and its projection on the brane. The particle oscillates along the extra dimension, and for a closed orbit on the brane, the corresponding trajectory in bulk is a closed wavy loop. }
\label{fig0}
\end{figure}

Without the extra dimension, the particle can move along a closed trajectory due to a potential field. Then, anytime the particle returns to the same point in its periodic motion, it will have the same amount of kinetic energy. In the presence of the extra dimension, regarding Eq.(\ref{1i}), the particle oscillates along the extra dimension. Let us assume that the particle returns to the same point on the brane, again due to the existence of a potential field, at time $t=T$. The equality $2\bar E_kdt=m\delta_{ij}\bar v^idx^j$, and Eq.(\ref{1i}) gives us the following form of $\psi$ at time $t=T$
\begin{equation}\label{1ii}
    \psi(T)=\frac{N}{(\bar E_k)^\frac{1}{4}}\sin(\frac{2}{\hslash}\int_0^T\bar E_kdt-\delta)=\frac{N}{(\bar E_k)^\frac{1}{4}}\sin(\frac{1}{\hslash}\oint\bar p_idx^i-\delta),
\end{equation}
where $\bar p_i=m\bar v_i$, and a closed line integration in the second equality is determined along a closed curve on 3-brane, see Fig. \ref{fig0}. Generally, after the particle returns to its initial point on the brane (initial coordinates), the extra dimension's value can differ from its initial value. Also, the particle's mass depends on the $\psi$, as we find in Eq.(\ref{sshh}). Thus, the mass will also change at the endpoint. One can raise the question of whether or not this change may vanish for certain orbits. 
As Fig.\ref{fig0} shows, the particle trajectory in the bulk space must be a closed wavy loop to have the same mass. In other words, $\psi(t_0)=\psi(t_0+T)$. This means
\begin{equation}\label{1j}
 \oint\bar p_idx^i=2\pi\hslash n,~~~n=1,2,3,...~,  
\end{equation}
which is the Bohr--Sommerfeld quantization rule. 
Although the particle's mass changes over time, its mass remains constant each time it reaches the same place. In other words, the particle's mass can vary at different places along its route, but the mass at the same point does not change over time.

The duration between two subsequent particle passages through the brane is $t=\pi/\omega$, where $\omega=2\bar E_k/\hslash$ is the angular frequency of oscillations along the extra dimension given by (\ref{Melin1b}). Therefore, Eq.(\ref{1j}) shows the emergence of the de Broglie wavelength, whence $\lambda=2\bar vt=2\pi\hslash/p$. As a result, brane gravity permits only a discrete series of motion states. In essence, this recalls the resonance property of de Broglie waves, i.e., the property that de Broglie used to reformulate the old Born--Sommerfeld quantum conditions for the first time. Furthermore, assume we take a position measurement on the brane in a specified direction.  As a result, the position measurement uncertainty is $ \Delta x_i\geq \frac{\lambda}{2}=\pi\hslash/\bar p_i$. At the same time, because the particle in the distance between these two points moves in bulk space, its momentum is uncertain. The average momentum of the particle between these two points is $m\cdot(\lambda/2)/t$. Therefore, the uncertainty in the measurement of  the momentum is given by $\Delta p_i\geq m\cdot(\lambda/2)/t=\bar p_i$. By multiplying these two uncertainties, we find
\begin{equation}
    \Delta x_i\Delta p_i\geq\pi\hslash,
\end{equation}
which is equivalent to the Heisenberg
inequality for conjugate position and momentum components of the particle.

As previously stated, the $4D$ trajectory of the particle on the brane is discrete, with the distance between sequence points equal to the particle's de Broglie wavelength. When the linear dimension of the system under study is significant compared to the de Broglie wavelength, the discretized behavior of the trajectory disappears, and the particle dynamics revert to conventional confined particle dynamics. Given that quantum systems have linear sizes similar to the particle's de Broglie wavelength, it is fair to expect that the unique behavior specified by Eqs.(\ref{1g:a}) and (\ref{1g:b}) will be evident in these systems. Also, in the same way, the minuscule de Broglie wavelength of the macroscopic objects prevents us from measuring their wave nature.
As a result, in the next chapter, we will look for examples of this kind in the equations mentioned above.

\section{Classic examples of quantum mechanics with new perspectives}

 Quantum mechanics enables the calculation of the properties and behavior of physical systems. Quantum mechanics limits the energy, momentum, angular momentum, and other properties of a bound system to discrete (quantized) values, in contrast to classical physics. It seems paradoxical, but quantum physics predicts that a particle can tunnel a potential barrier even though its kinetic energy is less than the potential's maximum.

We apply our non-relativist equations (\ref{1g:a}) and (\ref{1g:b}) to some quantum mechanics textbook examples before establishing the equivalency of particle motion along the extra dimension given by Eq.(\ref{1g:b}) with the SE. These simple examples provide us with a new perspective on quantum phenomena.

\subsection{Free particle}
Let us consider the motion of a free particle. Suppose the momentum of the particle at zero approximation is $\bar p_i$, the 3-velocity $\bar v^i=x^i/t$, and the corresponding kinetic energy is $\bar E_k=\bar p^2/2m$. Inserting this kinetic energy into Eq.(\ref{1g:b}) gives
\begin{equation}
    \label{1k}
    \psi=\psi_0\sin(\frac{2}{\hslash}\bar E_kt+\theta)=\psi_0\sin(\frac{1}{\hslash}\bar p_ix^i+\theta).
\end{equation}
The second expression has been obtained by replacing $2\bar E_kt=\bar p_i\bar v^it=\bar p_ix^i$. Also, one can introduce complex notations to distinguish the right-moving, $\exp(-i p_ix^i/\hslash)$, and left-moving, $\exp(i p_ix^i/\hslash)$ particles. The above solution shows that the particle oscillates along the extra dimension, and the particle appears in brane periodically with a wavelength of the oscillation equal to the de Broglie wavelength, $\lambda=2\pi\hslash/\bar p$.

\subsection{The infinite square well}
\begin{figure}[ht]
\centering
\includegraphics[width=10cm]{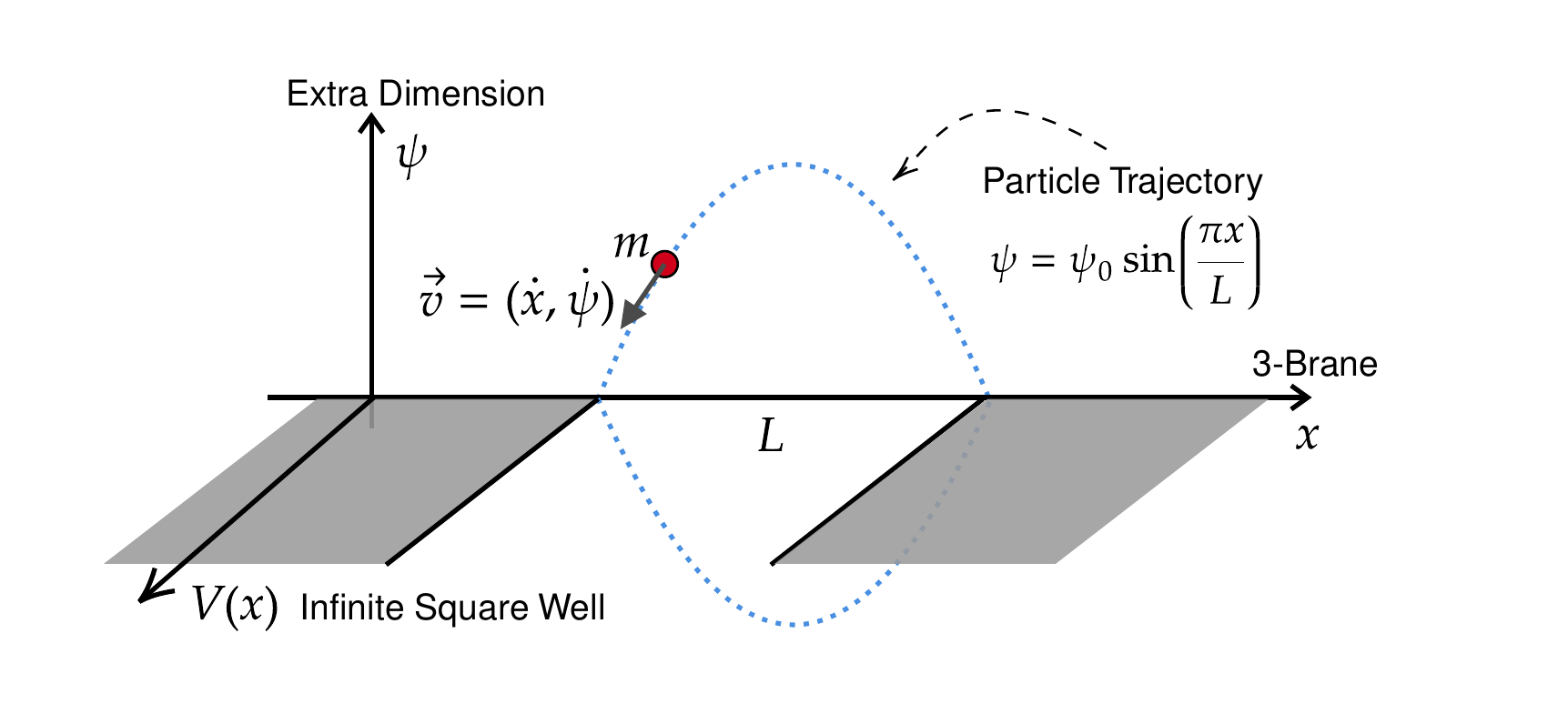}
\caption{A trajectory of a particle inside an infinite square well. It is semi-confined to the 3-brane and oscillates along the extra dimension. The figure shows the ground state of the particle, where $n=1$. }
\label{box}
\end{figure}

Consider a particle moves freely at $0<x<L$ and extra dimension; however, it is subject to an infinite repulsive force confined to the brane at the origin and $x=L$, see Fig.\ref{box}. 
Eq.(\ref{1g:a}) shows that the $x$-component of the particle's velocity equals a classical one. We have, $\bar p=m\bar v$, $t= x/\bar v$, and  $\bar E_kt= x\bar p/2$. Thus, the equation of motion (\ref{1g:b}) gives
\begin{equation}\label{1l}
    \psi=\psi_0\sin(\frac{2}{\hslash}\bar E_kt+\theta)=\psi_0\sin(\frac{\bar p x}{\hslash}+\theta).
\end{equation}
The particle cannot move in either region outside the well. Hence, $\theta=0$, and $\psi( x=0)=\psi( x=L)=0$. Inserting these boundary conditions into (\ref{1l}) lead us
\begin{equation}\label{1m}
\bar p=\frac{n\pi\hslash}{L},~~~~~\bar E_k=\frac{\bar p^2}{2m}=\frac{n^2\pi^2\hslash^2}{2mL^2},~~~~\psi=\psi_0\sin(\frac{n\pi x}{L}).
\end{equation}
The total energy (\ref{Energy}) of the particle is
\begin{equation}\label{1n}
  E=\bar E_k= \frac{n^2\pi^2\hslash^2}{2mL^2},~~~~~n=1,2,3,...~. 
\end{equation}

\subsection{The linear harmonic oscillator}
As a next example, let us consider a harmonic oscillator with a potential energy $V=\frac{1}{2}m\omega^2x^2$, where $m$ is the mass, and $\omega$ is its angular frequency. The confined movement of the particle at zero approximation is described by $x=x_0\sin(\omega t+\theta)$, and the particle's Kinetic energy is $\bar E_k=\bar E\cos^2(\omega t+\theta)$, where $\bar E$ is the mechanical energy, $\bar E=\bar E_k+V(x)$. Substituting this Kinetic energy into (\ref{1g:b}) results
\begin{equation}
 \frac{d^2\psi}{du^2}+\tan(u)\frac{d\psi}{du}+\left(\frac{2\bar E}{\hslash\omega}\right)^2\cos^4(u)\psi=0, 
 \end{equation}
where $u=\omega t$, and we assumed $\theta=0$. The solution of the above differential equation
gives $\psi$ in terms of the Kummer $M(\frac{3}{4}-\frac{\bar E}{2\hslash\omega},\frac{3}{2},\frac{2\bar E}{\hslash\omega}\sin^2(\omega t))$ and $U(\frac{3}{4}-\frac{\bar E}{2\hslash\omega},\frac{3}{2},\frac{2\bar E}{\hslash\omega}\sin^2(\omega t))$ functions. 
In order to have a semi-confined particle solution, we need to have a polynomial form for $\psi(t)$. Kummer functions diverge if the first index is not a negative integer, i.e., $\frac{3}{4}-\frac{\bar E}{2\hslash\omega}=-n$ or $\frac{3}{4}-\frac{\bar E}{2\hslash\omega}=(1-n)/2$. Thus, the oscillation is an orthogonal polynomial of the order $n\in \mathbb N$, obtained through the following relations
\begin{equation}
 M(-n,\frac{3}{2},z^2)=\frac{(-1)^nn!}{2(2n+1)!}\frac{H_{2n+1}(z)}{z},~~~~~~~
 U(-n,\frac{3}{2},z^2)=\frac{H_n(z)}{2^nz},
\end{equation}
where $H_n(z)$ are
the Hermite polynomials. One can summarize the above two solutions in one single formula 
\begin{equation}\label{1o}
    \psi(t)=Ne^{-\frac{\bar E}{\hslash\omega}\sin^2(\omega t)}H_n\left(\sqrt{\frac{2\bar E}{\hslash\omega}}\sin(\omega t)\right),~~~~~~~
    E=\hslash\omega\left(n+\frac{1}{2}\right),~~~~n=0,1,2,...~.
 \end{equation}   
Regarding $x=x_0\sin(\omega t)$,  the coordinate of the particle along the extra dimension, $\psi$, has the same function form for the wave function in QM for HS. In the following section, we show that Eq.(\ref{1g:b}) realizes the stationary SE.

\subsection{Bouncing Particle}
Consider a particle of mass $m$ bouncing vertically and elastically on a hard, reflective floor. The potential energy in a gravitational field (subject to a constant gravitational acceleration $g$) pointing in the $-x$ direction is $V(x)=mgx$. Assuming that the particle is falling down from some maximum height, say $\bar x(t=0)=x_\text{max}$ and $\bar v(t=0)=0$, then solutions of equations of motion for zero approximation are given by $\bar x(t)=x_\text{max}-\frac{1}{2}gt^2$, $\bar v(t)=-gt$.
The motion of an elastically reflected particle at the Earth's surface is periodic and confined to the interval $[0, x_\text{max}]$. The period of this motion is equal to $T =\sqrt{8x_\text{max}/g}$ and can be calculated using the condition $\bar x(T/2) = 0$. In maximum height, all kinetic energy $\bar E_k=mg^2t^2/2$ is transferred into
potential energy. Thus, the total energy of the particle is given by $E=mgx_\text{max}$. In this case, the equation of motion along the extra dimension, (\ref{1g:b}), becomes
\begin{equation}
    t\frac{d}{dt}\left(\frac{1}{t}\frac{d\psi(t)}{dt}\right)+\left(\frac{mg^2}{\hslash} \right)^2t^4\psi(t)=0.
\end{equation}
If we define $\eta=t^2$, the the above differential equation becomes
\begin{eqnarray}
    \frac{d^2\psi}{d\eta^2}+\left(\frac{mg^2}{2\hslash} \right)^2\eta\psi=0.
\end{eqnarray}
Regarding the boundary condition $\psi(\eta=0)=0$,
the following Airy function is the solution to the above equation 
\begin{equation}
    \psi(\eta)=\psi_0Ai\left(-\left(\frac{mg^2}{2\hslash} \right)^\frac{2}{3}\eta\right),
\end{equation}
where $\psi_0$ is a constant. Due to infinite gravitational potential, the particle is confined at the bouncing point, i.e.,  $\psi=0$. Therefore, for $t=\frac{T}{2}$ or $\eta=\frac{T^2}{4}=\frac{2E}{mg^2}$ we have
\begin{equation}
    Ai\left(-\frac{E}{mg}\left(\frac{2m^2g}{\hslash^2}\right)^\frac{1}{3}\right)=0.
\end{equation}
As we know, this is the condition for obtaining the eigenvalues of bouncing particles in QM.

\section{Derivation of the Schr\"odinger equation and tunneling}\label{Schro}
As the preceding examples demonstrate, there must be a close connection between particle dynamics in the extra dimension and the SE. This relationship is investigated in this section. After that, we will reinterpret the tunneling of particles through a potential barrier

\subsection{Schr\"odinger equation}

Let us start with a $1D$ problem where the particle's coordinate is $x(t)$, and it is influenced by a  potential $V(x)$. It may be more suitable to rewrite the equation of motion along the extra dimension (\ref{1g:b}) by eliminating time $t$, i.e., $\psi=\psi(x(t))$. Then, $d/dt=vd/dx\simeq \bar vd/dx+\mathcal O(\psi)$, and  equation (\ref{1g:b}) becomes
\begin{equation}\label{JS1}
   \frac{\hslash^2}{2m}\frac{d^2\psi}{dx^2}+(E-V)\psi=0,
\end{equation}
which is, in fact, the $1D$ stationary SE. In $3D$ case, consider a brane with local coordinates $(x^1,x^2,x^3)$,  $ds:=\sqrt{\delta_{ij}dx^idx^j}= v(t)dt$ be the differential element of the particle's trajectory in the 3-brane, and $\psi(t)=\psi(s(t))$. This means that we assumed that $\psi$ could be expressed as a function of a single variable $s$.  With these assumptions, Eq.(\ref{1g:b}) reduced to
\begin{equation}\label{Psi1}
   \frac{d^2\psi(s)}{ds^2}+\frac{2m}{\hslash^2}\bar E_k\psi(s)=0. \end{equation}
Then we have to apply Jeffery’s method \cite{1981AuJPh..34..113J,1982JPhA...15.2761L} to the above equation.  It is easy to verify that
\begin{equation}\label{Jef}
\partial_i\psi=\frac{d\psi}{ds}\partial_is,~~~~\delta^{ij}{\partial_i s}{\partial_j s}=1,~~~~\delta^{ij}{\partial_{ij}^2s}=0,~~~~~
\delta^{ij}{\partial^2_{ij}\psi}= \delta^{ij}{\partial_i s}{\partial_j s}\frac{d^2\psi}{ds^2} +\delta^{ij}{\partial_{ij}^2s}\frac{d\psi}{ds}.
\end{equation}
Inserting relations (\ref{Jef}) into (\ref{Psi1}) give us
\begin{equation}\label{psi3}
 \frac{\hslash^2}{2m} \delta^{ij}{\partial_i\partial_j}\psi+(E-V)\psi=0,  
\end{equation}
which is desired $3D$ SE.

 In QM, the complex-valued wave function associated with a quantum system is not itself considered a physical element of theory. Nevertheless, its absolute square, for instance, represents a probability distribution associated with particular outcomes of an experiment. In our model, the extra coordinate of the particle, which is the solution of  (\ref{1g:b}), is real-valued. However, by removing the time from (\ref{1g:b}) and rewriting it in terms of the other three coordinates of the particle, we found SE, which has complex solutions for $\psi$. To obtain the extra coordinate of the particle, we have to calculate the absolute value of $\psi$. While the above outline suggests numerous topics for future research, one obvious question concerns the status of the wave function in quantum mechanics and its relationship to motion along the extra dimension. While the wave function is a complex-valued entity, the particle's extra-dimensional oscillation is represented by a real-valued function. On the other hand, the observer and his measurement tools are confined to the brane. As a result, he cannot see the particle's $5D$ motion, which means he does not know the particle's direction of motion prior to the measurement process. As a result of his ignorance, he is forced to use a complex representation for motion along the extra dimension.

\subsection{Tunneling of particles through a potential barrier}
The phenomenon of a particle tunneling into a region where the potential energy function exceeds the particle's total energy is one of the most stunning pieces of evidence of the qualitative difference between quantum physics and classical limit. This would be impossible according to classical mechanics if we assumed space is $3D$.

Let us examine the 
physics of `one-dimensional' tunneling phenomena. In fact, according to our model, a $1D$ tunneling phenomenon is a $2D$ scattering problem: a particle moves towards the confined potential barrier to the brane, and it scatters along the extra dimension. Here, the simplest tunneling case over a $1D$ rectangular potential barrier, $V(x)=V_0\{\Theta(x)-\Theta(x-L)\}$ where $V_0>0$ and $\Theta(x)$ is the  Heaviside step function, will be considered. 
\begin{figure}[ht]
\centering
\includegraphics[width=10cm]{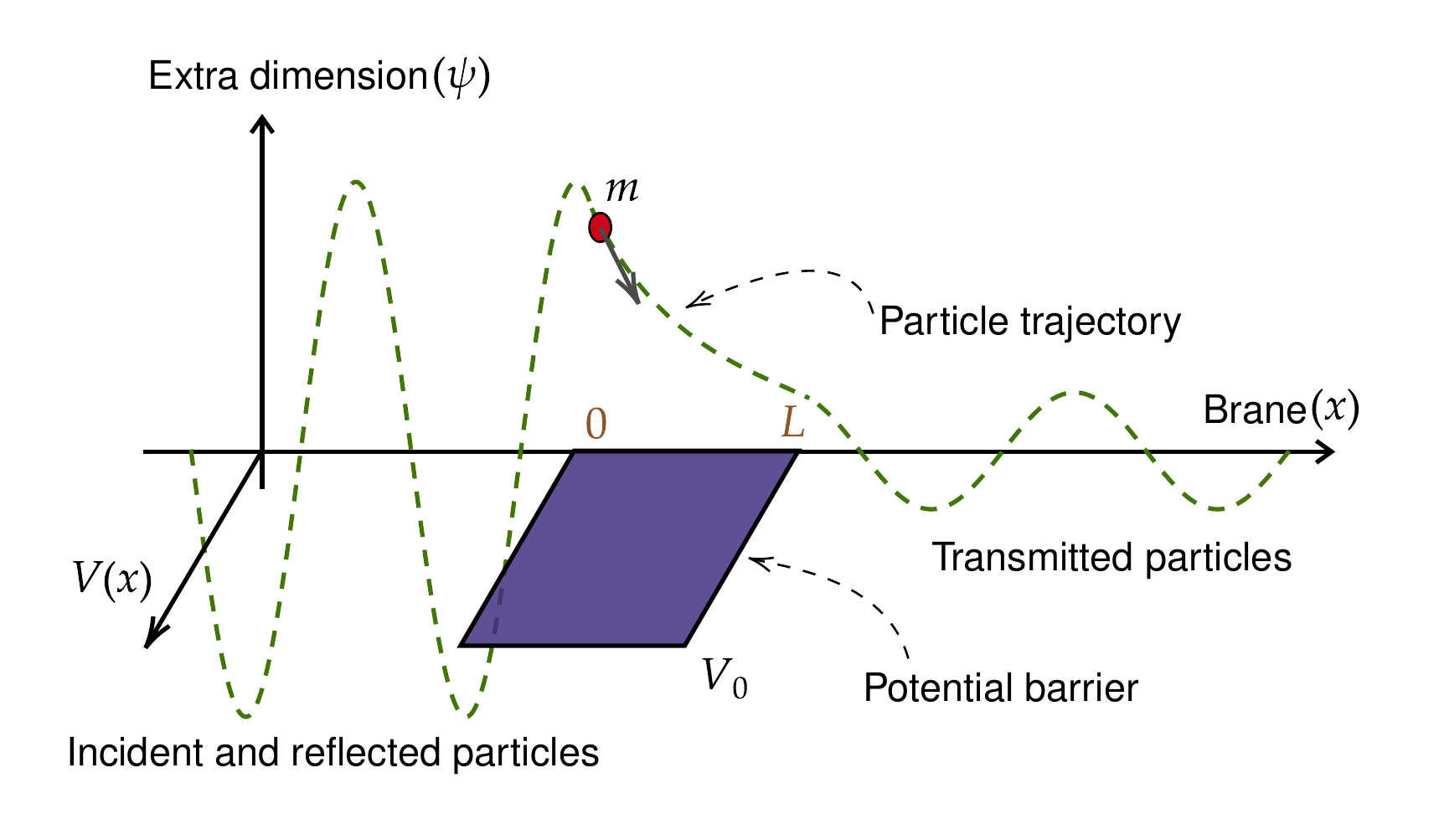}
\caption{A representation of the probable trajectory of a particle in the `tunneling' process.}
\label{fig2}
\end{figure}
Also, we will look at the scenario of $0< E< V_0$, when crossing through the barrier is conventionally prohibited.  
Let us assume an accelerator, in the left-hand side of the potential barrier of the Fig.\ref{fig2},  produces a beam of particles of mass $m$ at an early time $t_0$ with average momentum $\bar p$ (directed towards the target) and average energy $E=\bar p^2/(2m)$. Note that we just know the usual 3-components of the momentum of the particle, and our confined macroscopic measurement devices can not detect the extra component of the velocity of the particles. The projectiles will move toward the barrier, and as we saw in (\ref{1k}), they simultaneously will oscillate along the extra dimension, in which $\psi=\psi_0\sin(\frac{2}{\hslash}\bar E_kt+\theta)=\psi_0\sin(\bar px/\hslash+\theta)$. 
To distinguish the right mover particles from left moving particles, which all of which satisfy the earlier solution for $\psi$, we introduce complex notation  $A\exp(i\sqrt{2m\bar E}x/\hslash)$ and $B\exp(-i\sqrt{2m\bar E}x/\hslash)$  for right and left mover particles, respectively.
Thus for the left-hand side of the barrier, we have
\begin{equation}
 \psi=  Ae^{\frac{2i}{\hslash}\bar Et}+ 
  Be^{-\frac{2i}{\hslash}\bar Et}
= Ae^{i\sqrt{2m\bar E}x/\hslash}+ 
  Be^{-i\sqrt{2m\bar E}x/\hslash}.  
\end{equation}

The projectiles in the beam will be scattered by the potential, some of them will be reflected, and some of them will be transmitted over the barrier in the direction of the extra dimension to the right-hand side of the barrier, see Fig.\ref{fig2}. Regarding these notations, similar to the SE, the solution of (\ref{JS1}) for the barrier region will be
\begin{equation}
\psi=Ce^{\sqrt{2m(V_0-\bar E)}x/\hslash}+De^{-\sqrt{2m(V_0-\bar E)}x/\hslash}.
\end{equation}
In addition, on the right-hand side of the barrier, we have penetrated right mover particles, in which
\begin{equation}
\psi=Fe^{i\sqrt{2m\bar E}x/\hslash}. 
\end{equation}

\section{Conclusions}

Could it be that `classical' geometry is more fundamental than the rules of quantization, given the challenges in applying the laws of quantum physics to the geometry of space-time? In essence, this question is as old as quantum theory itself, and as noted in the introduction, many efforts have been made to provide an answer to it since modern physics' birth.
Gravitational interactions and phenomena are stated and interpreted in purely geometrical quantities in General Relativity. Matter fields are distributed over a geometrically curved space-time manifold, and test particles move on geodesics in this background. Apart from its remarkable phenomenological success, this attractive and intuitive geometric foundation is the cause for General Relativity's continuing appeal more than 100 years after its development. A positive response to the above question implies that the undeniable observable implications of quantization must derive from a more profound geometrical theory. This procedure will immediately encounter the determinism problem. This means that whereas uncertainties in measuring simultaneous canonical variables are just the result of unknown initial conditions in causal geometrical theories, they are an inevitable fundamental concept in conventional quantum mechanics.

The purpose of all of the models stated in the introduction (Feyn\`es--Nelson stochastic formulation,  scale relativity, Weylian geometry, Finsler geometry, and Kaluza--Klein  theory) is to derive the Schr\"odinger equation (or its relativistic extensions, such as the Klein--Gordon and Dirac equations) from generalized space-time geometry. Our model differs from prior models in that we obtained an entirely new equation (see Eq.(\ref{1g:b})) that reproduces the Schr\"odinger equation's spectrum results.
In addition, as shown by Eqs.(\ref{1b}) and (\ref{1h}), the existence of Newton's gravitational constant $G_N$, the cosmological constant $\Lambda$, and Planck's quantum of action $h=2\pi\hslash$ are all dependent on the tension of the brane $\lambda$. As shown by Eq.(\ref{1h}), $h$ is no longer a fundamental constant of nature, and its value may be computed in terms of other fundamental constants.

Last but not least, the wave function concept in our model is tied to the extra dimension(s). Quantum physics' most fundamental concept is the wave function. Initially, Schr\"odinger considered the wave function as a description of a real physical wave. However, this viewpoint was quickly supplanted by Born's probability interpretation, which has since become the accepted explanation of the wave function. Nevertheless, it is still unclear what the wave function is: is it just a representation of our restricted understanding of a system, or does it correspond to reality? According to no-go theorems, the wave function must be real if there is an objective reality. That conclusion, however, was based on dubious assumptions \cite{2015NatPh249R}.
Lundeen et al. provided in Ref.\cite{lundeen2011direct} a direct approach for measuring the real and imaginary parts of the wave function. Following that, there has been much interest in using weak values to determine a pure quantum state, see \cite{mirhosseini} and references therein. In the direction of giving a reality to wave function, we show that it denotes the particle's oscillation along the extra dimension. In this regard, measuring the wave function directly reckons the very existence of an extra dimension.

Even though the geometric formulation of  quantum mechanics described above is quite rudimentary, it reveals that classical brane gravity intrinsically contains the Schr\"odinger equation in the classical particle equations of motion. From the author's perspective, this is a crucial step toward developing a consistent causal explanation of quantum phenomena.

\appendix
\section{}\label{App}
Here, we show why in the geodesic equations for an electron (or a similar microscopic particle), we can omit the cosmological constant of the bulk space in the metric expansion (\ref{1f}).

One can rewrite the tangential components of the metric (\ref{1f}) in the following form
\begin{equation}
    \label{f1a}
    g_{\mu\nu}(x^\sigma,\psi)=g^{\alpha\beta}(x^\alpha,0)\Big\{g_{\alpha\mu}(x^\sigma,0)-\psi K_{\alpha\mu}\Big\}\Big\{g_{\beta\mu}(x^\sigma,0)-\psi K_{\beta\mu}\Big\}-\frac{1}{6}\Lambda_\text{B}g_{\mu\nu}(x^\sigma,0)\psi^2.
\end{equation}
The above form of $ g_{\mu\nu}(x^\sigma,\psi)$ explicitly exhibits two fundamental radii of the bulk space: The second term on the right-hand side of it establishes the presence of the curvature radius of the $5D$ bulk spacetime, $L_\text{(B)}=\sqrt{6/\Lambda_\text{B}}$. 
\begin{figure}[ht]
\centering
\includegraphics[width=10cm]{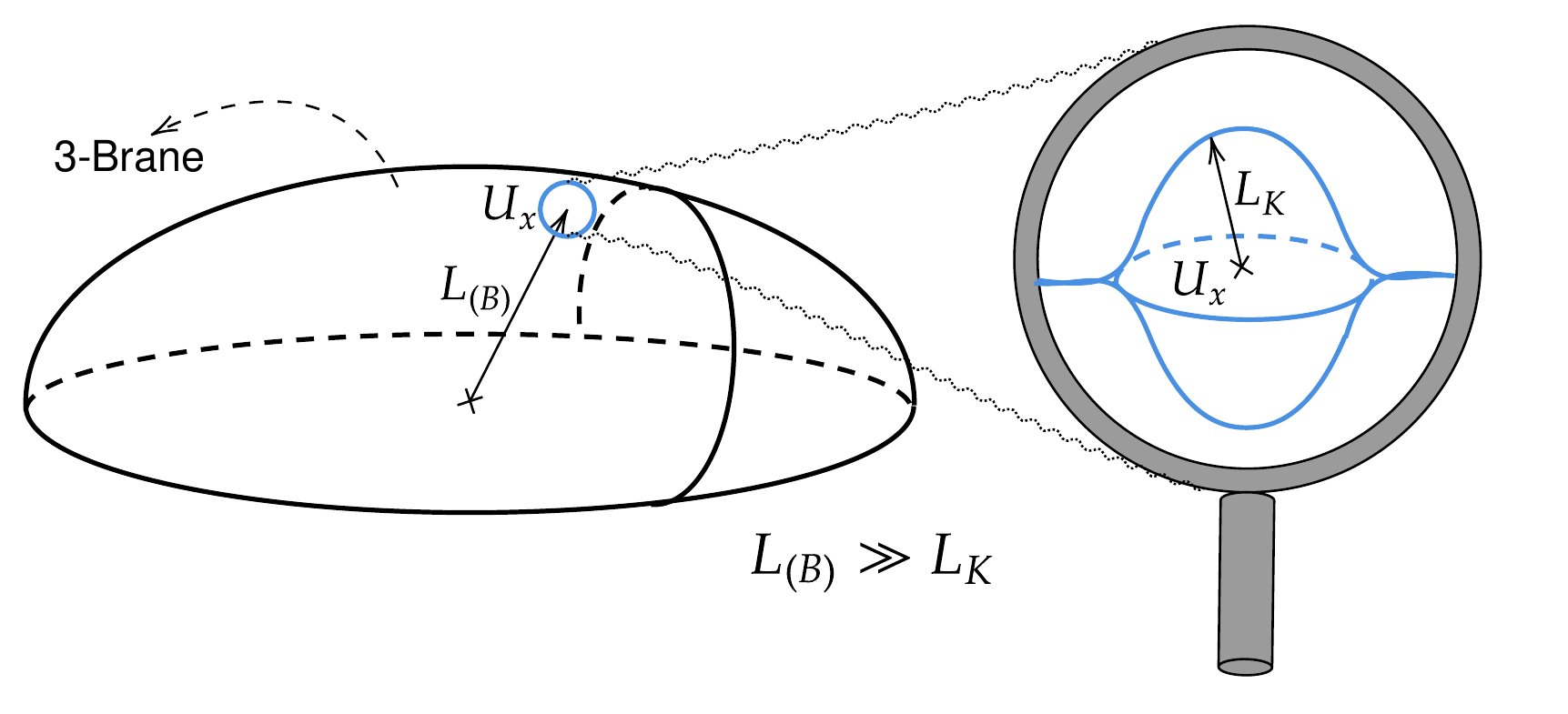}
\caption{The curvature radius of the bulk spacetime, $L_\text{(B)}$, and the extrinsic curvature radius $L_K$. Note that this figure is schematic. We have $Z_2$ symmetry throughout the brane, as we expected, and it does not realize mirror symmetry. }
\label{fig00}
\end{figure}
On the other, the form of the first term in the right-hand of (\ref{f1a})  shows that it becomes singular at the solutions of the following equation \cite{Eisenhart}
\begin{equation}
    \label{f2aa}
    \det(g_{\alpha\mu}(x^\sigma,0)-\psi K_{\alpha\mu})=0.
\end{equation}
This equation defines the extrinsic curvature radii, $L_K$, of 3-brane associated with the principal directions $\delta x^\mu$ \cite{Eisenhart} via
\begin{equation}
    \label{f3aa}
 (g_{\alpha\mu}(x^\sigma,0)-L_{(\mu)} K_{\alpha\mu})\delta x^\mu=0,  
\end{equation}
where $L_k$ is the smallest eigenvalue of the set $\{L_{(\mu)}\}$ \cite{G9}. 
According to Eq.(\ref{1b}), the extrinsic curvature is proportional to the distribution of matter fields on the 3-brane. In this work, we are concerned with the dynamics of non-relativistic microscopic particles, such as electrons, and the components of extrinsic curvature are provided by Eqs.(\ref{non}). As the result of the singularity of the first term, as Fig. \ref{fig00} shows, the presence of the particle itself at point $x^\alpha$ can restrict the particle's domain of motion along the extra dimension by $L_K$. Therefore, the motion along the extra dimension is restricted by two scales, $L_\text{(B)}$ and $L_K$. In fact, the ratio of the second term to the third term (in Eq.(\ref{1f}) with $\mathcal E_{\mu\nu}=0$) is approximately given by
\begin{equation}
    \label{f4aa}
    \frac{L_\text{(B)}^2}{L_K^2}\simeq\frac{|K_{\mu\beta}K^\beta_{~~\nu}\dot x^\mu\dot x^\nu|}{|\frac{1}{6}\Lambda_\text{B}g_{\mu\nu}\dot x^\mu\dot x^\nu|}=\frac{L_\text{(B)}^2}{\left(\frac{\hbar c}{2\bar E_k} \right)^2},
\end{equation}
where in the last equality, we used (\ref{non}) and (\ref{hslash}) to rewrite $K_{\mu\beta}K^\beta_{~~\nu}\dot x^\mu\dot x^\nu$ in terms of the kinetic energy of the particle and $\hbar$.
The value of $L_K$ for an electron approximately is $L_K\simeq \hbar c/m_ec^2\simeq10^{-13}$ m. The size of $L_\text{(B)}$ is model dependent. However, generally, it is bigger than $10^{-13}$ m and less than $5~\mu$m \cite{McWilliams:2009ym}. Therefore, one can neglect the effect of the bulk cosmological constant in geodesic equations for a microscopic particle like an electron. The particle is trapped inside a tiny geometric bubble (see Fig.\ref{fig00}) with the radius of its own Compton wavelength and oscillates inside it. On the other hand, for a graviton, the upper bound for its energy is $E_\text{graviton}\leq10^{-23}$ eV \cite{LIGOScientific:2019fpa}. In this case, we find $L_K\simeq 10^{16}$ m. This shows that, in contrast to the electron, in the case of gravitational waves (and neutrinos), the term $K_{\mu\beta}K^\beta_{~~\nu}$ is negligible in (\ref{1f}), and one has to keep the bulk cosmological constant in the geodesic equations (\ref{1d}). For macroscopic particles, the extrinsic curvature radius is quite modest. 
This, combined with the geodesic equation in the direction of the extra dimension, demonstrate that such particles oscillate along the extra dimension with very low amplitude and high frequency. To put it another way, the macroscopic particles are confined to the 3-brane.

The distribution of matter (including the particle and the rest of the matter fields) in the universe is the primary explanation for the quasi-confined motion of the particle along the extra dimension. As a result, even if we eliminate the particle's energy-momentum coupling, the rest of the matter in the universe will force the particle to oscillate in the extra dimension with gravitational pull. The particle will stay quasi-confined to the 3-brane even if we suppose that just the bulk cosmological constant exists and that the bulk space is de Sitter. Naturally, in this case, a large bulk cosmological constant is required for oscillations to be close to the 3-brane. Briefly, we can distinguish three scales in bulk space: the local extrinsic curvature radius, caused by coupling the particle energy-momentum tensor with the geometry of the brane (in which lambda plays the role of the coupling constant), the extrinsic curvature radii caused by the presence of matter fields in the brane, and finally, the curvature radius, caused by the bulk space cosmological constant. Each of the above possibilities can cause the particle to move in a quasi-confined manner, and each curvature radii listed can also restrict the amplitude of the particle's oscillation in the extra dimension.

It is worthwhile to note that several studies in the literature demonstrate that classical particles would be ejected from the brane if their motion were determined by geodesic equations in the Randall--Sundrum braneworld models (for an example, see \cite{Mueck:2000bb,Souza:2019jqz}). This occurs when the model is in its background state, that is, when the brane's induced metric is Minkowski, and the bulk space is a $5D$ anti-de Sitter. In the Randall--Sundrum braneworld models, the geometry itself does not offer the matter confinement (there exist no equilibrium points at the position of the brane, and therefore no confinement of particles purely due to gravitational
effects \cite{Dahia:2007vd}.); rather, a different process, such as the fermion-brane interaction postulated by V.A. Rubakov and M.E. Shaposhnikov in Ref.\cite{1983PhLB139R}, or a scalar field in the bulk space \cite{Dahia:2007ep,Gremm:1999pj} does so.
However, this is not a general result. For example, let us consider a warped product bulk space with a line element 
\begin{equation}
    \label{Ap1}
    ds^2=e^{2f(\psi)}g_{\mu\nu}(x,0)dx^\mu dx^\nu+d\psi^2.
\end{equation}
This is a spacial case of the metric (\ref{1a}), in which $g_{\mu\nu}(x,\psi)=e^{2f(\psi)}g_{\mu\nu}(x,0)$.
For the above simple warped bulk metric, the geodesic equations (\ref{1e}) turn to 
\begin{equation}\label{Ap2}
\begin{split}
\ddot x^\mu&+\Gamma^\mu_{\alpha\beta}\dot x^\alpha \dot x^\beta=-2\frac{df}{d\psi}\dot x^\mu\dot \psi+\frac{q}{mc}g^{\mu\nu}F_{\nu\beta}\dot x^\beta,\\
\ddot{\psi}&-\frac{df}{d\psi}g_{\mu\nu}(x,\psi)\dot x^\mu\dot x^\nu=0.
\end{split}
\end{equation}
As it is shown in Ref.\cite{Dahia:2007vd},  if $d^2f/d\psi^2<0$, the particles in this instance oscillate around the 3-brane. Indefinitely moving particles can now enter and exit the brane hypersurface (see Fig.\ref{fig0}) due to this quasi-confinement process. 
While we use a much more thorough measure in this piece, the same phenomenon occurs.
It should be noted that Randall--Sundrum brane models are not of relevance in this paper. But nevertheless, as demonstrated in reference \cite{2009EL8740006R}, the compatibility of the geodesic and geodesic division equations results in the Bohr--Sommerfeld quantization condition. Hence, quantization appears to be a typical tendency of brane models.

\section*{Declaration of competing interest}The author declares that he has no known competing financial interests or personal relationships that could have appeared to influence the work reported in this paper.

\section*{Data availability}
No data was used for the research described in the article.

\section*{Acknowledgements}

S.J. acknowledges financial support from the National Council for Scientific
and Technological Development -- CNPq, Grant no. 308131/2022-3.

\bibliographystyle{elsarticle-num}
\bibliography{main}

\end{document}